\documentclass[reprint,twocolumn, showpacs, aps,amssymb,amsmath]{revtex4-1}
\usepackage{geometry}                
\usepackage{graphicx}
\usepackage{amssymb}
\def\e{\begin{equation}}
\def\q{\end{equation}}
\def\m{\begin{eqnarray}}
\def\n{\end{eqnarray}}

\begin{document}
\title{Comment on the Identities of the Gluon Tree Amplitudes}

\author{S.-H. Henry Tye} 
\altaffiliation{Laboratory for Elementary Particle Physics, Cornell University, Ithaca, NY 14853, USA}

\author{and Yang Zhang}
\altaffiliation{Laboratory for Elementary Particle Physics, Cornell University, Ithaca, NY 14853, USA}



\begin{abstract} 
Recently, Bjerrum-Bohr, Damgaard, Feng and Sondergaard derived a set of new interesting quadratic identities of the Yang-Mills  tree scattering amplitudes. Here we comment that these quadratic identities of YM amplitudes actually follow directly from the KLT relation for graviton-dilaton-axion scattering amplitudes (in 4 dimensional spacetime). This clarifies their physical origin and also provides a simpler version of the new identities. We also comment that the recently discovered Bern-Carrasco-Johansson identities of YM helicity amplitudes can be verified by using (repeatedly) the Schouten identity. We also point out additional quadratic identities that can be written down from the KLT relations.
\end{abstract}
\maketitle

{\bf Introduction :}
Tree level scattering amplitudes have very compact forms which are not obvious at all if one follows the Feynman rules to evaluate the corresponding Feynman diagrams. Using a combination of symmetry, string theory techniques and the spinor helicity formalism, one can simplify the evaluation of these amplitudes considerably. In particular, the gluon (i.e., YM) amplitudes also obey a set of very useful identities. Until recently, all the known identities are linear identities :  such as the Kleiss-Kuijf relation \cite{Kleiss:1988ne} and the Bern-Carrasco-Johansson (BCJ) identities \cite{BCJ}. These identities also appear in loop diagrams \cite{BCJ,Bern:2010ue,Bern:2010yg}. Both can be easily proved \cite{BDV} using properties of string theory scattering amplitudes \cite{Plahte} \cite{KLT}. In fact, string theory improves the BCJ identities by (1) revealing more clearly its relation to the Jacobi identity \cite{Tye:2010dd}  and (2) reformulating them in terms of gauge invariant amplitudes instead of gauge-dependent quantities \cite{BDV}.  They also play a role in the KLT relation, which allows one to express graviton scattering amplitudes in terms of YM amplitudes (with their color factors stripped). 
Recently Bjerrum-Bohr, Damgaard, Feng and Sondergaard (BDFS) presented a field theory proof of the KLT relation and derived a new set of quadratic identities of YM amplitudes \cite{BDFS}. These identities are non-trivial, in the sense that they do not follow from the linear identities. They are very interesting in the sense that they are unexpected, at least from the YM theory point of view. 
 
 Here we like to point out that these quadratic identities actually follows directly from the KLT relation for the (massless) graviton-dilaton-axion scattering amplitudes. 
Recall that the KLT relation expresses the closed string tree scattering amplitudes in terms of the open string tree scattering amplitudes \cite{KLT}. At the zero-slope limit, the massless open string modes can include the YM fields while the closed string modes can include the graviton, the dilaton and the anti-symmetric tensor field, which is equivalent to an axion in 4 dimensional spacetime. So the graviton-dilaton-axion scattering amplitudes can be expressed as 
a sum of products of YM amplitudes. Now the dilaton and the axion combine to form a complex scalar field $\phi$, which has a conserved charge associated with it. Any charge-violating M-point graviton-dilaton-axion scattering amplitude must vanish, thus providing a quadratic identity of the M-point YM amplitudes.  

Since the same non-vanishing graviton amplitude can be expressed in terms of different sets of YM amplitudes in a variety of ways, we immediately obtain another set of quadratic identities. Although these quadratic identities follow immediately from the KLT relation, their usefulness has not been explored.

The recently discovered BCJ identities (including the Kleiss-Kuijf relations) are linear identities among the tree YM amplitudes. Here we would like to argue that the BCJ identities of YM helicity amplitudes can be proved by using the Schouten identity alone (besides kinematic identities). This is not surprising, since the BCJ identities can be expressed in a way that are dual to the Jacobi identity, and the Schouten identity has an analogous structure to the Jacobi identity. We like to emphasize that this does not negate the usefulness of the BCJ identities.

{\bf The BCJ relation and Schouten Identity :}
The Kleiss-Kuijf (KK) \cite{Kleiss:1988ne} and BCJ relations \cite{BCJ} for YM helicity amplitudes can be proved by using Schouten identity repeatedly, 
\begin{equation}
  \langle i j \rangle \langle k l \rangle =\langle i k \rangle \langle j l \rangle + \langle i l \rangle \langle k j \rangle
\end{equation}
Here we use the notation of \cite{Dixon:1996wi,Cachazo:2005ga}: for lightlike momentum $p_\mu$, $p_{a \dot a}=p_\mu \sigma^\mu_{a \dot a}=\lambda_a \tilde \lambda_{\dot a}$. The spinor products are defined to be $\langle \lambda ,\lambda'  \rangle =\epsilon_{ab} \lambda^a \lambda'^b$ and $[\tilde \lambda ,\tilde \lambda']=\epsilon_{\dot a \dot b} \tilde \lambda^{\dot a} \tilde \lambda'^{\dot b}$. Here $s_{ij}= (p_i+p_j)^2 = 2p_i \cdot p_j = \langle ij \rangle [ij]$.
Note the similarity between Schouten identity and Jacobi identity of Lie algebra.

To simplify the discussion, let us consider only the maximal helicity-violating (MHV) amplitudes here. Checking the 4- and 5-point cases is straightforward. For example, the KK relation for 4-point, the photon decoupling identity
\begin{equation}
  A(2^-1^-3^+4^+)+A(1^-2^-3^+4^+)+A(1^-3^+2^-4^+)=0
\end{equation}
which reads \cite{Parke:1986gb} \cite{Berends:1987me}, 
\begin{equation}
  \frac{\langle 12  \rangle^4 \big(\langle 23 \rangle \langle 41 \rangle+\langle 13 \rangle \langle 24 \rangle- \langle 12 \rangle \langle 34 \rangle\big)}{\langle 12 \rangle \langle 23 \rangle \langle 34 \rangle \langle 41 \rangle \langle 13 \rangle \langle 24 \rangle} =0         
\end{equation}
because of Schouten identity. The 4-point BCJ relation, $s_{12} A(2^-1^-3^+4^+)= s_{23} A(1^-3^+2^-4^+)$ is self-evident from the MHV amplitude expression.
The 5-point case is similar: the KK relation is, 
\begin{eqnarray}
  &&A(2^-1^-3^+4^+5^+)+A(1^-2^-3^+4^+5^+)+ \nonumber \\
 && A(1^-3^+2^-4^+5^+)+A(1^-3^+4^+2^-5^+)=0 
\end{eqnarray}
which reads,
\begin{eqnarray}
  &&\frac{\langle 12  \rangle^4 }{\langle 12 \rangle \langle 23 \rangle \langle 34 \rangle \langle 45 \rangle \langle 51 \rangle \langle 13 \rangle \langle 24 \rangle \langle 25 \rangle } \big(\langle 23 \rangle \langle 51 \rangle \langle 24\rangle  \nonumber \\
&&+\langle 13\rangle \langle 24\rangle \langle 25\rangle -\langle 12\rangle \langle 34\rangle \langle 25 \rangle 
  -\langle 12 \rangle \langle 23\rangle \langle 45\rangle \big)\nonumber \\
&& =0 
\end{eqnarray}
where the Schouten identity is used twice. The BCJ relation is,
\begin{eqnarray}
  && s_{12}A(2^-1^-3^+4^+5^+)-s_{23}A(1^-3^+2^-4^+5^+)\nonumber \\
&& -(s_{23}+s_{24})A(1^-3^+4^+2^-5^+)=0
\end{eqnarray}
whose left hand side is proportional to,
\begin{eqnarray}
  &&\langle 32 \rangle \langle 24 \rangle \langle 51\rangle \langle 21 \rangle [21] - \langle 21\rangle \langle 34\rangle \langle 52\rangle\langle 23\rangle [23] \nonumber \\
 &-&\langle 21\rangle \langle 45\rangle \langle 32 \rangle\langle 32 \rangle [32] -\langle 21 \rangle \langle 45\rangle \langle 32\rangle \langle 24\rangle [24] \nonumber \\
  &=& \langle 32 \rangle \langle 21 \rangle  \langle 24 \rangle \big(\langle 51\rangle [21]+\langle 53\rangle [23]+\langle 54\rangle [24]\big)\nonumber \\
  &=&0
\end{eqnarray}
where the first step uses the Schouten identity and the last step follows from momentum conservation : $\sum_{j=1}^M p_j^{\mu}=0 \rightarrow \sum_j   \langle kj\rangle [jl] =0$ and $\langle ii\rangle=0$ and  $[ii]=0$.
This feature can be generalized to the $M$-point MHV amplitudes. For example, consider a BCJ identity for $M$-points \cite{BCJ} \cite{BDV} \cite{Tye:2010dd},
\begin{equation}
  \sum_{i=3}^{M} \bigg(\sum_{j=3}^i s_{2j}\bigg) A (13...i,2,(i+1)...M) = 0
\label{BCJ}
\end{equation}
where the helicities are $(1^-2^-3^+....M^+)$ and the label $j=M+1$ should be identified with $j=1$. The left hand side of (\ref{BCJ}) reads,
 \begin{eqnarray}
  &&\sum_{j=3}^M s_{2j} \bigg( \sum_{i=j}^M  A (13...i,2,(i+1)...M)\bigg) \nonumber \\
&=& \frac{\langle 12 \rangle^4}{\langle 13 \rangle \langle 34 \rangle ... \langle M 1\rangle }\sum_{j=3}^M s_{2j} \bigg(\sum_{i=j}^M \frac{\langle i, i+1\rangle}{\langle i2\rangle \langle 2,i+1\rangle}\bigg)\nonumber \\
&& \label{Parke-Taylor}
\end{eqnarray}
The sum over $i$ can be calculated by using Schouten identity in $(M-j)$ steps, say,
\begin{eqnarray}
  && \frac{\langle j, j+1\rangle}{\langle j,2\rangle \langle 2, j+1\rangle }+\frac{\langle j+1,j+2\rangle }{\langle j+1,2 \rangle \langle 2,j+2\rangle}=\frac{\langle j, j+2\rangle }{\langle j,2 \rangle \langle 2,j+2\rangle}\nonumber \\
&&  
\end{eqnarray}
and etc. The final result is,
\begin{equation}
  \sum_{i=j}^M \frac{\langle i, i+1\rangle}{\langle i2\rangle \langle 2,i+1\rangle}=\frac{\langle j,1\rangle}{\langle j,2\rangle \langle 21\rangle}.
\end{equation}
Therefore, the left hand side of (\ref{BCJ}) reads,
\begin{eqnarray}
&& \frac{\langle 12 \rangle^4}{\langle 13 \rangle \langle 34 \rangle ... \langle M 1\rangle } \sum_{j=3}^M  \frac{s_{2j} \langle j,1\rangle}{\langle j,2\rangle \langle 21\rangle}\nonumber \\
&\propto& \sum_{j=3}^M \frac{\langle j,1\rangle [2,j]}{\langle 21 \rangle }=0, 
\end{eqnarray}
where we have used momentum conservation.

{\bf The Quadratic Identity :}
The massless sector of a closed string theory contains the graviton, the dilaton and the antisymmetric tensor field $B_{\mu \nu}$. In 4-dimensional spacetime, $B_{\mu \nu}$ has only one degree of freedom, and one may identify it as an axion $a$ : $\partial^{\mu} a = \epsilon^{\mu \nu \rho \sigma} \partial_{\nu} B_{\rho \sigma}$. 
To discuss the 4-dimensional polarization tensor structure for the graviton, the dilaton and the axion, we use the light cone gauge: consider a massless particle and the polarization vector or tensor which has only transverse components. A polarization tensor can be decomposed as,
\begin{eqnarray}
  e_{ij}=\frac{e_{ij}+e_{ji}-\delta_{ij} e_{kk}}{2}+\frac{e_{ij}-e_{ji}}{2}+\frac{\delta_{ij} e_{kk}}{2}\nonumber 
\\
\end{eqnarray}
where $i=1,2$ labels the two transverse directions. The three terms correspond to the graviton, the axion (the antisymmetric tensor field) and the dilaton modes. We may choose the positive polarization vector $\epsilon^+=(1,i)$ while the negative polarization $\epsilon^-=(1,-i)$. The addition of two spin 1 (with polarizations $\epsilon$ and ${\tilde \epsilon}$) is straightforward. It is easy to see that a graviton has polarization mode $\epsilon^+ {\tilde \epsilon}^+$ or $\epsilon^- {\tilde \epsilon}^-$, an axion has polarization $\epsilon^+ {\tilde \epsilon}^-- \epsilon^-{\tilde \epsilon}^+$ and the dilaton has polarization $\epsilon^+ {\tilde \epsilon}^-+\epsilon^- {\tilde \epsilon}^+$. The dilaton and the axion combine to form a massless complex scalar field $\phi$, which has a global conserved charge associated with it. All scattering (tree or loop) amplitudes must obey this charge conservation.


Within helicity amplitudes, graviton $j$ has helicity $\epsilon_j^{\pm} {\tilde \epsilon}_{\tilde j}^{\pm}=j^{\pm} {\tilde j}^{\pm}$ and an incoming positively charged scalar field $j$ may be identified with helicity  $j^{+} {\tilde j}^{-}$ while an incoming negatively charged scalar field may be identified with helicity $j^{-} {\tilde j}^{+}$.  Any charge conservation-violating amplitude ${\mathcal A}$ must vanish. That is, any amplitude with unequal numbers of positively and negatively charged scalar fields will vanish. Let us start with a non-vanishing $M$-graviton scattering amplitude. Following the BDFS notation, let $n_+$ ($n_-$) be the number of $"+"$ ($"-"$) helicities in YM amplitude $A$ that have been flipped in YM amplitude $\tilde A$. Then the resulting amplitude vanishes whenever $n_+ \ne n_-$. 

Let us consider the 4-point case to establish some notation : the graviton-dilaton-axion scattering amplitude takes the form
\e
{\mathcal A}_4 = - s_{12} A(1234) {\tilde A}(2134) 
\label{a4}
\q
where both $A$ and $\tilde A$ are YM amplitudes.
For 4-graviton amplitudes, helicity conservation requires 2 with helicity $(++)$ and 2 with helicity $(--)$. So the only non-vanishing amplitude has the form
\e
{\mathcal A}_4 = - s_{12} A(1^-2^-3^+4^+) {\tilde A}(2^-1^-3^+4^+) 
\q
Note that both $A$ and $\tilde A$ are maximal helicity-violating amplitudes. For $(n_+, n_-)=(1,1)$, say  
\e
{\mathcal A}_4 =  - s_{12} A(1^-2^-3^+4^+) {\tilde A}(2^+1^-3^-4^+)  
\q
the amplitude describes the graviton-$\phi$ scattering. For  $(n_+, n_-)=(2,2)$, ${\mathcal A}_4$ describes the $\phi$-$\phi$ scattering. For $n_+-n_- \ne 0$, ${\mathcal A}_4=0$ because the charge conservation is violated. Mathematically, we see that it vanishes because $\tilde A=0$. This case does not yield a quadratic identity. Rather it yields the well known results ${\tilde A}(1^+2^-3^+4^+)=0$ and ${\tilde A}(1^+2^+3^+4^+)=0$ without having to look into the detailed structure of these amplitudes.

Next consider the 5-graviton scattering case,
\begin{eqnarray}
&&{\mathcal A}_5 =  s_{12}s_{34} A(1^-2^-3^+4^+5^+) {\tilde A}(2^-1^-4^+3^+5^+) \nonumber \\
&+& s_{13}s_{24} A(1^-3^+2^-4^+5^+) {\tilde A}(3^+1^-4^+2^-5^+)
\label{a5}
\end{eqnarray}
For $n_+-n_- \ne 0$, the resulting ${\mathcal A}_5=0$. For example, for $(n_+, n_-)=(1,0)$, we have
\begin{eqnarray}
&0&=s_{12}s_{34} A(1^-2^-3^+4^+5^+) {\tilde A}(2^-1^-4^+3^-5^+)+ \nonumber \\ 
&&s_{13}s_{24} A(1^-3^+2^-4^+5^+) {\tilde A}(3^-1^-4^+2^-5^+)
\end{eqnarray}
It is easy to verify this quadratic identity by using the explicit formulae for the MHV amplitudes. 
For $M \ge 6$, non-MHV amplitudes appear in the quadratic identities.

Following from the KLT relation, each identity contains $(M-3)![\frac{1}{2}(M-3)]! [\frac{1}{2}(M-3)]!$ ($M \text{odd}$) 
or $(M-3)![\frac{1}{2}(M-4)]! [\frac{1}{2}(M-2)]!$ ($M \text{even}$) terms quadratic in the M-point YM amplitudes. So the number of terms in each identity is substantially less than $[(M-3)!]^2$ : 2 vs 4  for $M=5$, 12 vs 36 for $M=6$, 96 vs 576 for $M=7$ etc.. Of course, one may use the BCJ identity (repeatedly) to reduce the number of terms in the quadratic identities to that in the KLT relations. 

{\bf Additional Quadratic Identities :} Although ordering is important in YM amplitudes $A_M$ and ${\tilde A}_M$, ${\mathcal A}_M$ is invariant. For example, besides Eq.(\ref{a4}), one can express the same ${\mathcal A}_4$ in other forms,
$
{\mathcal A}_4= - s_{13} A(2134) {\tilde A}(1324) = . . . .
 $
Comparing this expression to Eq.(\ref{a4}), one obtains a quadratic identity, 
$$s_{12} A(1234) {\tilde A}(2134) - s_{13} A(2134) {\tilde A}(1324) =0.$$
These identities are valid for any choice of helicities. 
Since the same ${\mathcal A}_M$ can be expressed in terms of $A_M$ and ${\tilde A}_M$ in many different ways, we obtain a large set of new quadratic identities by equating any 2 different expressions. This result is known but not emphasized in earlier work. Presumably these identities can be proved by using the BCJ identity (repeatedly); still their usefulness may follow from the relative ease in writing them down.

For example, expressing the same ${\mathcal A}_6$ in 2 different ways \cite{KLT}, we end up with a quadratic identity
\begin{eqnarray}
&& A(123456)[s_{35}  {\tilde A}(215346)+ s_{3(45)} {\tilde A}(215436)]\nonumber \\
 &-& {\tilde A}(123456)[s_{13} {A}(231546)+ s_{3(12)} A(321546)]\nonumber \\
 &+& {\text{permutations of }}  (234) =0 
 \label{ab6}
\end{eqnarray}
where $s_{3(45)}=s_{34}+s_{35}$.
Here, in each case where ${\mathcal A}_6=0$, we get a set of quadratic identities instead of only one. For example, besides
\m
&&[s_{35}  {\tilde A}(2^-1^-5^+3^+4^+6^+)+ s_{3(45)} {\tilde A}(2^-1^-5^+4^+3^+6^+)]\nonumber \\
 &\times&A(1^-2^-3^-4^+5^+6^+)+ {\text{permutations of }}  (234) =0 \nonumber
\n
we also get
\m
&&[s_{13} {A}(2^-3^-1^-5^+4^+6^+)+ s_{3(12)} A(3^-2^-1^-5^+4^+6^+)] \nonumber \\
 &\times& {\tilde A}(1^-2^-3^+4^+5^+6^+) +{\text{permutations of }}  (234) =0 \nonumber
\n
We leave it for the reader to write down other inequivalent identities coming from the vanishing of the same ${\mathcal A}_6$.

{\bf Acknowledgments :} We thank Poul Damgaard for remarks on this comment.
This work is supported by the National Science Foundation under grant PHY-0355005.

\end{document}